\documentclass[a4, 11pt]{article}

\usepackage[T1]{fontenc}

\usepackage{amssymb}        
\usepackage{latexsym}
\usepackage{array}

\usepackage{graphicx}

\author{}

\date{}
\usepackage{algorithm, algorithmic}

\title{Characterisation of Strongly Stable Matchings}

\author{Pratik Ghosal and Adam Kunysz and Katarzyna Paluch \\
University of Wroc{\l}aw}

\newcommand{\dowod}{\noindent{\bf Proof.~}}
\newcommand{\koniec}{\hfill $\Box$\\[.1ex]}
\newtheorem{fact}{Fact}

\newtheorem{lemma}{Lemma}
\newtheorem{theorem}{Theorem}

\newtheorem{definition}{Definition}
\newtheorem{observation}{Observation}

\newtheorem{property}{Property}

\begin{document}

\maketitle

\thispagestyle{empty}

\begin{abstract}
An instance of a strongly stable matching problem (SSMP) is an undirected bipartite graph $G=(A \cup B, E)$, with   an adjacency list of each vertex being a linearly ordered list of ties, which are subsets of vertices
equally good for a given vertex. Ties are disjoint and may contain one vertex. A matching $M$ is a set of vertex-disjoint edges. An edge $(x,y) \in E \setminus M$ is a {\em blocking edge} for $M$ if $x$ is either unmatched or strictly prefers $y$ to its current partner in $M$, and $y$ is either unmatched or strictly prefers $x$ to its current partner in $M$ or is indifferent between them. A matching is {\em strongly stable} if there is no blocking edge with respect to it. We present an algorithm for the generation of all strongly stable matchings, thus solving an open problem already stated in the book by Gusfield and Irving \cite{GI}.  It has previously been shown that strongly stable matchings form a distributive lattice and although the number of strongly stable matchings can be exponential in the number of vertices, we show that there exists a partial order with $O(m)$ elements representing all strongly stable matchings, where $m$ denotes the number of edges in the graph.  We give two  algorithms that construct two such representations: one  in $O(nm^2)$ time and the other in $O(nm)$ time, where $n$ denotes the number of vertices in the graph. Note that the construction of the second representation has the same time complexity as that of computing a single strongly stable matching.
\end{abstract}

\section{Introduction}
An instance of a strongly stable matching problem (SSMP) is an undirected bipartite graph $G=(A \cup B, E)$, with   an adjacency list of each vertex being a linearly ordered list of ties, which are subsets of vertices
equally good for a given vertex. Ties are disjoint and may contain one vertex. Thus if vertices $b_1$ and $b_2$ are neighbours of $a$ in the graph $G$, then either (1) $a$ (strictly) prefers $b_1$ to $b_2$, which we denote as   $b_1 \succ_{a} b_2$; or (2) $b_1$ and $b_2$ are in a tie on an adjacency list of $a$, and then we say that $a$ is indifferent between $a_1$ and $a_2$ and denote it as $b_1 =_a b_2$; or (3) $a$ (strictly) prefers $b_2$ to $b_1$. If a vertex $a$ prefers $b_1$ to $b_2$ or is indifferent between them, we say that $a$ weakly prefers $b_1$ to $b_2$ and denote as $b_1 \succeq_{a} b_2$. A matching $M$ is a set of edges, no two of which share an endpoint. Let $e=(v,w)$ be an edge contained in a matching $M$. Then we say that vertices $v$ and $w$ are matched in $M$ and that $v$ is a partner of $w$ in $M$, which we also denote as $v=M(w)$. If a vertex $v$ has no edge of $M$ incident to it, then we say that $v$ is free or unmatched in $M$. An edge $(x,y) \in E \setminus M$ is a {\em blocking edge} for $M$ if $x$ is either unmatched or strictly prefers $y$ to its current partner in $M$, and $y$ is either unmatched or weakly prefers $x$ to its current partner. In other words, an edge $(x,y)$ is blocking with respect to $M$, if by getting matched to each other, neither of the vertices $x$ and $y$ would become worse off and at least one of them would become better off than in $M$. A matching is {\em strongly stable} if there is no blocking edge with respect to it.

As is customary, we call the vertices of the graph men -- those belonging to the set $A$, and women -- the ones belonging to $B$. An ordered adjacency list of a vertex $v$ is also called its preference list and denoted $L_v$.
The problem of computing a strongly stable matching, if it exists, has already been solved. Let $n$ and $m$ denote the number of, correspondingly, vertices and edges in the graph. Irving \cite{Irving94} gave an $O(n^4)$ algorithm for computing strongly stable matchings for instances in which the bipartite graph is complete and there are equal number of men and women. In \cite{1999Manlove} Manlove extended the algorithm to incomplete bipartite graphs; the extended algorithm has running time $O(m^2)$. In \cite{KMMP} Kavitha, Mehlhorn, Michail and Paluch gave an $O(nm)$ algorithm for SSMP.

In this paper we study the problem of characterising the set of all strongly stable matchings. The problem was already stated in 1989 in the book by Gusfield and Irving \cite{GI} as one of the $12$ open problems and posed again in many subsequent papers and also in a recent book by Manlove \cite{Manbook}. Let us mention here that in contrast to the problem of characterisation of the set of all strongly stable matchings, the structure of the set of all stable matchings in the stable matching problem, which is the classical  variant without ties, is well understood. The set of stable matchings forms a lattice  and although the number of stable matchings may be exponential, there are known  compact representations of all stable matchings that can be constructed in $O(m^2)$ or even $O(m)$ time. 

The set of strongly stable matchings has also been shown to form a distributive lattice \cite{Manlove02}.  However, no characterisation of such a set has been  known so far. We give two compact representations of the set of all strongly stable matchings that can be constructed in, correspondingly, $O(nm^2)$ and $O(nm)$ time, where $n$ and $m$ denote the number of vertices and edges in the graph. We also show how to efficiently construct a partial order  on the elements of the
representation. The presented compact representations as well as the posets on the sets of elements of these representations can be used to solve a number of problems connected with strongly stable matchings. In particular, we are able to efficiently enumerate all strongly stable matchings, we can compute all {\em stable pairs}, where a pair $(a,b)$ is stable if $e=(a,b) \in E$ and there exists a strongly stable matching containing $e$, and many others.
Also, any known algorithm for computing a strongly stable matching outputs either a man-optimal strongly stable matching or woman-optimal strongly stable matching. A man-optimal strongly stable matching has the property that each man is matched in it to the best partner he can have in any strongly stable matching. A woman-optimal strongly stable matching has an analogous property. It has been conjectured by Feder \cite{Feder} that it may be NP-hard to decide if there exists a strongly stable matching which is neither man-optimal or woman-optimal. In the paper we disprove this conjecture.

One of the two representations of the set of stable matchings consists of $O(m)$ elements, each of which is a man-optimal  stable matching among the set of all  stable matchings containing a given edge. In the case of the strongly stable matchings we give an analogous representation, which for any stable pair $(a,b)$ has a class  of strongly stable matchings, which are man-optimal among the set of strongly stable matchings containings $(a,b)$. We show that computing such a class can be reduced to computing a man-optimal strongly stable matching in an appropriately constructed instance of SSMP. The  reduction is surprisingly simple. It is described in Section \ref{Preliminaries}. The second representation of the set of stable matchings can be derived from differences (so called {\em rotations}) between consecutive matchings in  a maximal sequence of stable matchings beginning with a man-optimal stable matching and ending with a woman-optimal stable matching. Our second representation can be analogously obtained from differences between consecutive classes of matchings in a maximal sequence of classes of strongly stable matchings.  
This second representation can be constructed in $O(nm)$ time -- note that the  time equals the running time  of the algorithm computing a single strongly stable matching.

\section{Preliminaries} \label{Preliminaries}

In this section we recall some well-known theorems and theory concerning strongly stable matchings.
We are going to make use of the following two theorems:

\begin{theorem}\cite{KMMP} \label{KMMP}
There is an $O(nm)$ algorithm to determine a man-optimal strongly stable matching of the given instance or report that no such matching exists.
\end{theorem}

\begin{theorem}\cite{IwamaMMM99} \label{iwama}
In a given instance of SSMP,   the same  vertices are matched   in all strongly stable matchings. 
\end{theorem}

We introduce some notation and  definitions.

For a given edge $(m, w)$ any matching $M$ such that $(m, w) \in M$ is called an {\em $(m, w)$-matching}.
Let us denote the set of all strongly stable matchings of $G$ by 
$\mathcal{M}_G$. Let $\mathcal{M}_G(m, w)$ be the set of strongly stable $(m, w)$-matchings in $G$.

Let $L_v$ be a preference list of vertex $v$. $L_v$ is a linearly ordered list of ties. The first tie on $L_v$ contains highest ranked edges for $v$ and we say that the {\em rank} of every edge $(v,w)$ contained in this list is $1$ and denote $rank_v(v,w)=1$. Similarly, the second tie on $L_v$ contains edges of rank $2$ with respect to $v$ and so on. For  a strongly stable matching $M$, by $rank_M(v)$ we donote $rank_v(v,M(v))$.

We define an equivalence relation $\sim$ on $\mathcal{M}_G$ as follows. 
\begin{definition}
For two strongly stable matchings $M$ and $N$, $M \sim N$ if and only if each man $m$ is indifferent between $M(m)$ and $N(m)$. 
Denote by $[M]$ the equivalence class containing $M$, and denote by $\mathcal{X}$ the set of equivalence classes of $\mathcal{M}_G$ under $\sim$.
\end{definition}

For two strongly stable matchings $M$ and $N$, we say that {\em $M$ dominates $N$} and write $M \succeq N$ if each man $m$ weakly prefers $M(m)$ to $N(m)$. If $M$ dominates $N$ and there exists a man $m$ who prefers $M(m)$ to $N(m)$, then we say that {\em $M$ strictly  dominates $N$}
and we call $N$ a {\em successor} of $M$.

Next we define a partial order $\preceq^{*}$ on $\mathcal{X}$:
\begin{definition}
 For any two equivalence classes $[M]$ and $[N]$, $[M] \preceq^{*} [N]$ if and only if $M \preceq N$.
\end{definition}

Let $M$ and $N$ be any two matchings. Then the symmetric difference $M \oplus N = (M \setminus N) \cup (N \setminus M)$ contains {\em alternating paths} and {\em alternating cycles}, where  a path $p$ (cycle $c$) is {\em alternating (with respect to matching $M$)} if its edges alternately belong to $M$ and to $E\setminus M$. If $M$ and $N$ are two strongly stable matchings of the same graph, then by Theorem \ref{iwama},
$M \oplus N$ contains only alternating cycles. Alternating cycles of $M\oplus N$ display an interesting property captured in:

\begin{lemma} \cite{Manlove02} \label{Manlove1} \label{cykle} Let $M$ and $N$ be two strongly stable matchings.  Consider any alternating cycle $C$ of $M \oplus N$. Let $(m_0, w_0, m_1, w_1, ..., m_{k - 1}, w_{k - 1})$ be a sequence of vertices of $C$ where $m_i$ are men and $w_i$ are women. Then there are only three possibilities:

\begin{itemize}
	\item $(\forall m_i) w_i =_{m_i} w_{i + 1}$ and $(\forall w_i) m_i =_{w_i} m_{i - 1}$
	\item $(\forall m_i) w_i \prec_{m_i} w_{i + 1}$ and $(\forall w_i) m_i \succ_{w_i} m_{i - 1}$
	\item $(\forall m_i) w_i \succ_{m_i} w_{i + 1}$ and $(\forall w_i) m_i \prec_{w_i} m_{i - 1}$
\end{itemize}

Subscripts are taken modulo $k$.
\end{lemma}

Below we introduce two operations transforming pairs of strongly stable matchings into other strongly stable matchings.

\begin{definition}
Let $M$ and $N$ be two strongly stable matchings. Consider any man $m$ and his partners $M(m)$ and $N(m)$. 

By $M \wedge N$ we denote the matching such that:

\begin{itemize}
	\item if $M(m) \succeq_{m} N(m)$ then $(m, M(m)) \in M \wedge N$
	\item if $M(m) \prec_{m} N(m)$ then $(m, N(m)) \in M \wedge N$
\end{itemize}

Similarly by $M \vee N$ we denote the matching such that:

\begin{itemize}
	\item if $M(m) \succ_{m} N(m)$ then $(m, N(m)) \in M \vee N$
	\item if $M(m) \preceq_{m} N(m)$ then $(m, M(m)) \in M \vee N$
\end{itemize}

\end{definition}

From  \cite{Manlove02}  it follows that both $M \vee N$ and $M \wedge N$ are strongly stable matchings, and 
$M, N \preceq M \vee N$ and $M, N \succeq M \wedge N$. 

We extend operations $\vee$ and $\wedge$ to the set $\mathcal{X}$ of equivalence classes. Let $[M], [N] \in \mathcal{X}$. Denote $[M] \vee [N] = [M \vee N]$, $[M] \wedge [N] = [M \wedge N]$.

\begin{theorem}\cite{Manlove02}
The partial order $(\mathcal{X}, \preceq^{*})$ with operations meet $\vee$ and join $\wedge$  defined above forms a distributive lattice.
\end{theorem}

Note that the set $\mathcal{M}_G(m, w)$ is closed under meet and join operations. It implies that the set of equivalence classes of $\mathcal{M}_G(m, w)$ under $\sim$ forms a sublattice of $\mathcal{X}$. hence there is a single man optimal equivalence class of $\mathcal{M}_G(m, w)$.

\section{Construction of the auxiliary graph $G_{(m,w)}$} \label{trick}

In this section we describe an $O(nm)$ algorithm for computing a man-optimal matching in $\mathcal{M}_G(m, w)$ or deciding that no such matching exists.

Let $(m, w) \in E$ be an edge of the graph $G$. The idea is very simple. In order to calculate a man-optimal matching in $\mathcal{M}_G(m, w)$ we are going to define a new graph $G_{(m,w)}$, such that there is a one-to-one correspondence between strongly stable matchings in $\mathcal{M}_{G_{(m, w)}}$ and $\mathcal{M}_G(m, w)$. Moreover, a man-optimal matching of $G_{(m,w)}$ is going to be a man-optimal matching of $\mathcal{M}_G(m, w)$.

Let $G_{(m,w)} = (A \cup B, E')$ be a subgraph of $G$. Preference lists of $G_{(m,w)}$ are derived from preference lists of $G$. Below we describe which edges should be removed from $E$ in order to obtain the set $E'$.

\begin{itemize}
	\item $(m, w)$ is removed from $E$
	\item let $m'$ be a vertex such that $(m', w) \in E$ and $m \succ_{w} m'$. We remove $(m', w)$ from $E$.
	\item let $m'$ be a vertex such that $(m', w) \in E$ and $m =_{w} m'$. We remove $(m', w)$ from $E$. Additionally we remove every edge $(m', w') \in E$ such that $w \succ_{m'} w'$.
	\item let $m'$ be a vertex such that $(m', w) \in E$ and $m \prec_{w} m'$. We remove $(m', w)$ from $E$. Additionally we remove every edge $(m', w') \in E$ such that $w \succeq_{m'} w'$.

	\item let $w'$ be a vertex such that $(m, w') \in E$ and $w \succ_{m} w'$. We remove $(m, w')$ from $E$.
	\item let $w'$ be a vertex such that $(m, w') \in E$ and $w =_{m} w'$. We remove $(m, w')$ from $E$. Additionally we remove every edge $(m', w') \in E$ such that $m \succ_{w'} m'$.
	\item let $w'$ be a vertex such that $(m, w') \in E$ and $w' \succ_{m} w$. We remove $(m, w')$ from $E$. Additionally we remove every edge $(m', w') \in E$ such that $m \succeq_{w'} m'$.
\end{itemize}

This concludes the construction of the graph $G_{(m,w)}$.

\begin{lemma}
Let $M \in \mathcal{M}_G(m, w)$. Then $M' = M \setminus \{(m, w)\} \in \mathcal{M}_{G_{(m,w)}}$.
\end{lemma}

\dowod
We will prove that $M' \subseteq E'$ and that $M'$ is  a strongly stable matching of $G_{(m,w)}$.

To prove $M' \subseteq E'$ we need to observe that none of the removed edges ($E \setminus E'$) is matched in $M$. Let us assume by contradiction that an edge $(m', w')$ was removed from $E$ and is matched in $M$. Obviously  $m \neq m'$ and $w \neq w'$. From the construction of $G_{(m, w)}$ it follows that there is an edge $(m, w')$ or $(m', w)$ which caused the removal of $(m', w')$. We can easily check that such an edge blocks $M$. It leads to a contradiction.

Strong stability of $M'$ is straightforward -- if there were an edge $e$ blocking $M'$, it would also block $M$. \koniec

\begin{lemma}
Let $M'$ be some strongly stable matching of $G_{(m,w)}$. If $M' \cup \{(m, w)\}$ is a strongly stable matching of $G$, then for each strongly stable matching $N$ of $G_{(m,w)}$, matching $N \cup \{(m, w)\}$ is a strongly stable matching of $G$. If $M' \cup \{(m, w)\}$ is not a strongly stable matching of $G$, then $\mathcal{M}_G(m, w) = \emptyset$.
\end{lemma}

\dowod

Let $M'$ be any strongly stable matching of $G_{(m, w)}$. Consider an edge $(m, w)$. Denote $M = M' \cup \{(m, w)\}$. From the construction of $G_{(m, w)}$ it follows that only edges from the set $E \setminus E'$ can potentially block $M$. We will prove that if any edge blocks $M$, then set $\mathcal{M}_G(m, w)$ is empty. Let us analyse the construction of the graph $G_{(m, w)}$. We have the following cases:

\begin{itemize}
	\item let $m'$ be a vertex such that $(m', w) \in E$ and $m \succ_{w} m'$. We removed $(m', w)$ from $E$.
	\item let $m'$ be a vertex such that $(m', w) \in E$ and $m =_{w} m'$. We removed $(m', w)$ from $E$. Additionally we removed every edge $(m', w') \in E$ such that $w \succ_{m'} w'$.
	\item let $m'$ be a vertex such that $(m', w) \in E$ and $m \prec_{w} m'$. We remove $(m', w)$ from $E$. Additionally we remove every edge $(m', w') \in E$ such that $w \succeq_{m'} w'$.	
	\item let $w'$ be a vertex such that $(m, w') \in E$ and $w \succ_{m} w'$. We removed $(m, w')$ from $E$.
	\item let $w'$ be a vertex such that $(m, w') \in E$ and $w =_{m} w'$. We removed $(m, w')$ from $E$. Additionally we removed every edge $(m', w') \in E$ such that $m \succ_{w'} m'$.

	\item let $w'$ be a vertex such that $(m, w') \in E$ and $w \succ_{m} w'$. We removed $(m, w')$ from $E$. Additionally we removed every edge $(m', w') \in E$ such that $m \succeq_{w'} m'$.
\end{itemize}

Case 1. An edge $(m', w)$ cannot block $M$.

Case 2. Note that from the construction of $G'$, if vertex $m'$ is matched in $M$, then neither $(m', w)$ nor $(m', w')$ can block $M$. If vertex $m'$ is unmatched in $M$, then from Theorem \ref{iwama} vertex $m'$ is unmatched in every strongly stable matching of $G'$. Let us assume that there exists some matching $N \in \mathcal{M}_G(m, w)$. Then $N' = N \setminus{(m, w)}$ is strongly stable in $G'$, so $m'$ is unmatched in $N'$. Hence $(m', w)$ blocks $N$, contradiction. 

We omit proofs of the remaining cases as these proofs are analogous to the proof of Case 2. \koniec

From above lemma, we conclude that either every strongly stable matching of $\mathcal{M}_{G_{(m, w)}}$ corresponds to some strongly stable matching of $\mathcal{M}_G(m, w)$, or none of them. Additionally we can easily compute a man-optimal strongly stable matching $M'$ in $G_{(m, w)}$ and check if $M' \cup \{(m, w)\}$ is strongly stable in $G$.
This implies the following theorem:

\begin{theorem}
Let $(m, w) \in E$. There is an $O(nm)$ algorithm for deciding whether $\mathcal{M}_{G}(m, w)$ is empty, and computing a man-optimal matching of $\mathcal{M}_{G}(m, w)$ if it exists.
\end{theorem}

\section{Basic representation} \label{basic}

In this section we prove the existence of a compact representation $I(\mathcal{M}_G)$ of the lattice $\mathcal{M}_G$. This representation is a generalization of the representation given in \cite{GI} for the classical stable marriage problem. Representation $I(\mathcal{M}_G)$ is simple to construct and  its correctness is easy to prove.  However, its construction takes $O(nm^2)$ time. 

Recall that equivalence classes of $\mathcal{M}_G(m, w)$ under $\sim$ form a sublattice of $\mathcal{M}_G$. Thus $\mathcal{M}_G(m, w)$ contains its own equivalence class of man-optimal strongly stable matchings.

\begin{definition}
By $M(m, w)$ we denote the equivalence class of man-optimal strongly stable $(m, w)$-matchings.
\end{definition}

\begin{definition}
An equivalence class of a matching $N$ is called irreducible if $[N]_{\sim} = M(m, w)$ for some $m,w$.
\end{definition}

By $I(\mathcal{M}_G)$ we denote the set of irreducible equivalence classes. We will consider $(I(\mathcal{M}_G), \preceq)$ as the partial order with the dominance relation inherited from $\mathcal{M}_G$.

A subset $S$ of $I(\mathcal{M}_G)$ is said to be closed in $I(\mathcal{M}_G)$ if there is no element in $I(\mathcal{M}_G) \setminus S$ that precedes an element in $S$.

Let $S \subseteq I(\mathcal{M}_G)$ be a closed set. Denote $\bigvee S = \bigvee_{T \in S}T$. Obviously $\bigvee S$ is an equivalence class of $\sim$. Hence every closed subset of $(I(\mathcal{M}_G), \preceq)$ corresponds to an equivalence class. We will prove that it is a bijection from the set of closed subsets of $(I(\mathcal{M}_G), \preceq)$ to the set of equivalence classes of $\sim$.

\begin{definition}
Let $M$ be any strongly stable matching. We define the irreducible support $U(M)$ to be 
$$U(M) = \{ M(m, w) : (m, w) \in M \}$$
\end{definition}

\begin{lemma} \label{lem1}
Let $M$ be any strongly stable matching. Then $[M] = \bigvee U(M)$.
\end{lemma}\dowod
Suppose that $[M] \neq \bigvee U(M)$. There is a man $m_1$ such that $(m_1, w_1) \in M$ and for every matching $M' \in \bigvee U(M)$, it holds that $w_1 \neq _{m_1} M'(m_1)$.
Note that $M(m_1, w_1)$ is in $U(M)$, so  $M'(m_1) \prec_{m_1} w_1$. There must be a pair $(m_2, w_2) \in M$ such that in any matching from the class $M(m_2, w_2)$ man $m_1$ gets matched to  a woman strictly worse than $w_1$. Class $M(m_2, w_2)$ dominates  class $[M]$, because $(m_2, w_2) \in M$. This gives a contradiction because $m_1$ prefers $w_1$ to any partner of any matching in $M(m_2, w_2)$. 
\koniec

By $C(U(M))$ we denote the set of all irreducible matchings that dominate some matching in $U(M)$, i.e. $C(U(M))$ is the closure of $U(M)$.

\begin{lemma} \label{lem2}
Let $M$ be a strongly stable matching. Then $[M]_{\sim} = \bigvee C(U(M))$.
\end{lemma} 
\dowod If $[M] \preceq^{*} [N]$ then $[M] \vee [N] = [N]$, so $\bigvee C(U(M)) = \bigvee U(M)$ since each matching in $\bigvee C(U(M))$ dominates some matching in $\bigvee U(M)$.\koniec

\begin{lemma} \label{lem3}
Let $M$ be a strongly stable matching. $[M] = \bigvee S$ for a set $S$ of equivalence classes that excludes $[M]$ if and only if $[M] \notin I(\mathcal{M}_G)$
\end{lemma}
\dowod $(\Leftarrow)$ follows from  Lemma \ref{lem2}.

$(\Rightarrow)$ If $[M] = \bigvee S$, then every class in $S$ dominates $[M]$. So if $[M] \notin S$, then for any pair $(m, w) \in M$ there is a class $[M'] \in S$, such that $M(m) =_m M'(m)$ and $M'$ strictly dominates $M$. Hence $[M] \neq M(m, w)$ and $[M]$ cannot be an irreducible class.\koniec

\begin{lemma} \label{lem4}
If $S$ and $T$ are distinct closed subsets of $I(\mathcal{M}_G)$, then $\bigvee S \neq \bigvee T$.
\end{lemma}
\dowod Since $S$ and $T$ are closed and $S \neq T$ one of the maximal matchings of $S \cup T$ (with respect to dominance) cannot be in $S \cap T$. So one of the sets (say $S$ without loss of generality) contains a class $[M]$ that does not dominate any matching in $T$. Moreover for some $m$ and $w$ we have that $[M] = M(m, w)$. Since $[M] \in S$, vertex $m$ has a partner no better than $w$ in any matching in $\bigvee S$. We claim that $m$ has a better partner than $w$ in every matching in $T$, so $\bigvee S \neq \bigvee T$. 

To prove this fact, suppose that $(m, w') \in M'$ and $w' \preceq_m w$ for some $[M'] \in T$. From the definition of $M(m, w)$ there is a matching $N \in M(m, w)$ such that $(m, w) \in N$. We can easily see that $N \wedge M'$ contains $(m, w)$, so $M(m, w) \preceq M(m, w) \wedge [M'] \preceq [M']$. 

Thus $M(m, w)$ dominates $M'$ contradicting the fact that $M(m, w)$ dominates no matching in $T$. \koniec

\begin{lemma}
If S is a closed subset of $I(\mathcal{M}_G)$ and $[M]_{\sim} = \bigvee S$, then $S = C(U(M))$.
\end{lemma}

The following theorem is an immediate consequence of Lemmas 4, 5, 6, 7.

\begin{theorem}
The function $S \rightarrow \bigvee S$ is a bijection between the nonempty closed subsets of $I(\mathcal{M}_G)$ and $\mathcal{X}$.
\end{theorem}

\begin{theorem} \label{run2}
Representation $(I(\mathcal{M}_G), \prec)$ can be constructed in time $O(nm^2)$. 
\end{theorem}

\dowod It is easy to see that the set $I(\mathcal{M}_G)$ can be computed in time $O(nm^2)$. It suffices to run the algorithm described in Theorem 9 for each edge $(m, w) \in E$. Obviously the set $I(\mathcal{M}_G)$ has at most $m$ elements. In order to determine the precedence relation on $(I(\mathcal{M}_G), \prec)$  we simply examine each pair of equivalence classes of $I(\mathcal{M}_G)$ and test whether one class dominates the other one. Each test clearly takes $O(n)$ time. This shows that the construction takes $O(nm^2)$ time. \koniec

\section{A Maximal Sequence of Strongly Stable Matchings}

In this section we will be interested in computing a sequence of strongly stable matchings $M_0 \succ M_1 \succ \ldots \succ M_z$ such that $M_0$ is a man-optimal strongly stable matching, $M_z$ is a woman-optimal strongly stable matching and for each $1 \leq i \leq z$, there exists no strongly stable matching $M'$ such that $M_{i-1} \succ M' \succ M_i$. We will call such a sequence - {\em a maximal sequence of strongly stable matchings}. In order to do this, we need to be able to compute a strict successor of any strongly stable matching $M$, where by a {\em strict successor} of $M$ we mean any strongly stable matching $M'$, which is a successor of $M$, i.e., $M \succ M'$ and such that there exists no strongly stable matching $M''$ such that $M \succ M'' \succ M'$.

Let $M$ be a strongly stable matching $M$  and $m$ a vertex in $A$. Suppose that there exists a strongly stable matching $M'$ such that $m$ gets a worse partner in $M'$ than in $M$, i.e.,  $M(m) \succ_m M'(m)=w'$. What edge incident to $m$ can potentially belong to $M'$? 
Obviously it must be an edge $(m,w)$ such that $M(m) \succ_m w$. By Lemma \ref{cykle}, we also get  that $m \succ_w M(w)$.
This way we get that any edge $(m,w)$ such that  $w \prec_m M(m) \ \wedge \  M(w) \prec_w m$ potentially belongs to a strict successor $N$ of $M$ such that $m$ has a worse partner in $N$ than in $M$. In the algorithm computing a strict successor of $M$, the set $E_c$ contains  for each man $m$ highest ranked edges incident to him  that potentially belong to some strict successor $N$ of $M$  such that $M(m) \succ_m N(m)$.

We can observe that if man $m$ gets a worse partner in a strongly stable matching $M'$, then it automatically means that certain other men must also get worse partners in $M'$ and certain women must get better partners in $M'$. For example, let us assume that $M'(m)=w' \prec_m w=M(m)$. Then, if there exists $w_1$ such that $w_1 =_m w$ and $m=_{w_1} M(w_1)$, then $w_1$ must have a better partner in $M'$ than in $M$, (otherwise $(m,w_1)$ would block $M'$) 
and as a result, by Lemma \ref{cykle}, $w_1'$s current partner $M(w_1)$ must have a worse partner in $M'$ than in $M$. Similarly, if there exists
$w_1$ such that (1) $w_1 =_m w'$ and $m \succ_{w_1} M(w_1)$ or (2)  $w \succ_m w_1 \succ_m w'$ and $m =_{w_1} M(w_1)$, then $w_1$ must have a better partner in $M'$ and $M(w_1)$ must have a worse partner in $M'$.

In Algorithm given below we maintain a directed  graph $G_d=(V, E_d)$, whose every edge $(m,w) \in E_d \cap M$ is directed from $w$ to $m$ and every other edge $(m,w)$ is directed from $m$ to $w$.  $G_d$ satisfies:

\begin{property} \label{prop}
Let  $M$ be a currently considered strongly stable matching and $x$ a vertex such that    $rank_M(x) \neq rank_{M_z}(x)$. Then graph $G_d$ constructed with respect to $M$ has the property that for every vertex $y$ reachable from $x$ in $G_d$ and any strongly stable matching $N$ such that  $M \succ N$ and $rank_M(x) \neq rank_{N}(x)$ it holds $rank_M(y) \neq rank_{N}(y)$.
\end{property}

A strongly connected component $S$ of a directed graph $G'=(V',E')$ is such a maximal set of vertices $S \subseteq V'$  that for every pair of vertices $x,y \in S$ vertex $y$ is reachable from $x$, i.e., there exists a directed path from $x$ to $y$ visiting only vertices of $S$. We say that $e=(v,w)$ is an outgoing edge of $S$ if $v \in S$ and $w \notin S$. The number of outgoing edges of $S$ is denoted as $outdeg(S)$.
Depending on the context, we treat a strongly connected component $S$  as a set of vertices, a set of (undirected) edges or a directed subgraph.
We say that a matching $M$ is {\em perfect on $S$} if every vertex of $S$ is matched in $M \cap S$.

We can notice that  strongly connected components of $G_d$ help in finding strict successors of the considered strongly stable matching in the following sense:

\begin{observation}
Let $M$ be a strongly stable matching and $N$ its  successor. Then the set $X=\{v: rank_M(v) \neq rank_N(v)\}$ has the property that  each strongly connected component $S$  of $G_d$ is either a subset of $X$ or is disjoint with $X$. Also, $X$ has no outgoing edge in $G_d$.
\end{observation}

In the algorithm while computing a strict successor of a  given strongly stable matching $M$ we consider each strongly connected component $S$ of $G_d$ with $outdeg(S)=0$ and try to find a perfect matching on $S$ in the graph $G_c$.
If we are successful, then we prove that this gives us a strict successor of $M$. Otherwise, we change the graphs $G_c$ and  $G_d$ by allowing edges of lower rank and continue.

Another graph, which we keep in Algorithm is $G_c$. We will prove that it satisfies:
\begin{property} \label{pro2}
Let $m$ be any man and $N$ any  strict successor of $M$ such that $M(m) \succ_m N(m)$. Then $rank_N(m) \geq  min\{rank_m(m,v) : (m,v) \in E' \cup E_c\}$. 
\end{property}

\begin{figure}[h]
{\em Algorithm }
\begin{algorithmic}[1]
\STATE let $M_0$ be any man-optimal strongly stable matching of $G$
\STATE let $M_z$ be any woman-optimal strongly stable matching of $G$
\STATE $M \leftarrow M_0$
\STATE let $M'$ contain edge $(m,M(m))$ for every man $m$ such that $M(m)=_m M_z(m)$ % otherwise $m$ is free in $M'$
\STATE let $E_d$ contain all edges of $M$ 
\STATE let $G_d$ be a directed graph $(V, E_d)$ in which every edge $(m,w) \in E_d \cap M$ is directed from $w$ to $m$ and every other edge $(m,w)$ is directed from $m$ to $w$
\STATE $E' \leftarrow E \setminus E_d$
\STATE let $E_c= M'$ and let $G_c=(V,E_c)$
\STATE for a vertex $x$ let $S(x)$ denote a strongly connected component of $G_d$ containing $x$
\STATE for each $(m,w) \in M$ remove from $E'$ each edge $(m',w)$ dominated by $(m,w)$ and each edge $(m, w')$ such that $w' \succ_m w$
\STATE let $i = 1$ 
\STATE set phase number $j=1$
%\STATE initialize graphs $G_d = (A \cup B, E_d)$, $G_c = (A \cup B, \emptyset)$
\REPEAT

	\WHILE {$(\exists  m \in A$) (  $deg_{G_c}(m) = 0$ and $outdeg(S(m))=0$)}
		\STATE add the set $E_m$ of top choices of $m$ from  $E'$ to $E_d$
		 \IF {$outdeg(S(m)=0$} 
		         \STATE add every edge $(m,w) \in E_m$ such that $m \succ_w M(m)$ and $M(m) \succ_m w$ to $E_c$
		         \STATE for every edge $(m,w)$ of $E_c$ that becomes dominated by some newly added edge $(m', w)$  remove it from $G_c$ 	
						 \STATE remove $E_m$ from $E'$
		 \ENDIF 
	\ENDWHILE
	
	\WHILE {$(\exists m \in A$) ($m$ is free in $M'$ and $outdeg(S(m))=0$)}
	        \IF {an alternating path from $m$ to a free woman  $w$ in $E_c$ exists} 
					\STATE let $w$ be a free woman in $M'$ reachable from $m$  by an alternating path $p$ in $E_c$
					\STATE $M' \leftarrow M' \oplus p$
					\ELSE \STATE let $Z$ be the set of men reachable from $m$ by alternating paths in $E_c$
					      \STATE let $N(Z)$ be the women adjacent to $Z$ in $E_c$
								\STATE delete all lowest ranked edges in $E_c \cup E'$  incident to any $w \in N(Z)$	% E_c i E'?
					\ENDIF			
	\ENDWHILE

	\WHILE {$(\exists S) (outdeg(S)=0$ and ($M'$ perfect on $S$)}
	   \STATE $M \leftarrow (M' \cap S) \cup (M\setminus S)$
	   \STATE $M_i \leftarrow M$
	   \STATE  output $M_i$
	   \STATE $i\leftarrow i+1$
	   \STATE $M' \leftarrow M' \setminus S$
	   \STATE update $G_c$ and $G_d$: $E_c \cap S$ contains only edges of the form $(m,M(m))$ such that $m$ is a man and $M(m)=_m M_z(m)$;
		 an edge $(m,w)$ stays in $G_d$ only if $rank_m(w)= rank_M(m)$ and $rank_w(m) \leq rank_M(w)$. 
	\ENDWHILE
	
	\STATE $j \leftarrow j+1$
\UNTIL{ $(\forall v \in A)$ $rank_{M}(v) = rank_{M_{z}(v)} $}

%%%%%%%%%%%%%%%%%%%%%%%%%%%%%%%%%%%%%%%%%%%%%%%%%%%%%%%%%%%%%%%%%%%%%%%%%%%%%%%
\end{algorithmic}
\caption{Algorithm for computing a maximal sequence of strongly stable matchings}
\label{algo}
\end{figure}

\subsection{Correctness of Algorithm}

Below we prove the correctness  of Algorithm computing a maximal sequence of strongly stable matchings. We begin with the following simple observations.
\begin{fact} \label{F}
During the whole execution:
\begin{enumerate}
\item $E_c \subseteq E_d$  

\item Let $l(m)= min\{rank_m(m,v) : (m,v) \in E' \cup E_c\}$. Then every edge $e=(m,w)$ of $E_d$ satisfies $l(m) \geq rank_m(e) \geq rank_M(m)$ and $rank_w(e) \leq rank_M(w)$.

\item Each edge $(m,w)$ of $E_c$ is contained in some strongly connected component $S$ of $G_d$ with $outdeg(S)=0$.
\end{enumerate}
\end{fact}
\dowod
The second point follows 15 and 18 of Algorithm.
The third point follows from lines 16-18 of Algorithm. \koniec

\begin{fact}
If we show, that we never delete an edge $e$ of $E_c$ which belongs to a strongly stable matching $N$ dominated by the current matching $M$, then it implies
that Property \ref{pro2} is satisfied. 
\end{fact}

\begin{lemma}
Assuming that at some point Algorithm  satisfies Property \ref{pro2}, it also satisfies Property $\ref{prop}$.
\end{lemma}
\dowod Suppose that at some point of the execution Property \ref{pro2} is satisfied. Let $M$ be a current strongly stable matching, whose strict successor we want to compute and $m$ any man such that $M(m) \succ_m M_z(m)$. Let $N$ be any  strict successor of $M$ such that $M(m) \succ_m N(m)$. 
By Fact \ref{F}(2) any edge $e=(m,w)$ of $E_d \setminus M$ satisfies $l(m) \geq rank_m(e) \geq rank_M(m)$ and $rank_w(e) \leq rank_M(w)$. Edge $e$ is directed in $G_d$ from $m$ to $w$. We want to show, that if $rank_N(m) \neq rank_M(m)$, then $rank_N(w) \neq rank_M(w)$.  By Property \ref{pro2} $rank_N(m) \geq  l(m)$.  This means that $rank_N(w) < rank_M(w)$. Otherwise $e$ blocks $N$ or Lemma \ref{cykle} does not hold. 

If $e=(m,w)$ is an edge of $E_d \cap M$, then by Lemma \ref{cykle}, if $rank_N(w) \neq rank_M(w)$, then $rank_N(m) \neq rank_M(m)$.

Thus, we have shown, that for every edge $(x,y)$ of $G_d$ and any strongly stable matching $N$ dominated by $M$, it holds $rank_M(x) \neq rank_{N}(x)$ implies $rank_M(y) \neq rank_{N}(y)$. Therefore lemma is proved.
\koniec

\begin{lemma} \label{lem5}
No edge $e$ deleted in line 19 of Algorithm can belong to any 
strongly stable matching $N$  dominated by $M$. 
\end{lemma}
\dowod
Suppose that the algorithm wants to delete an edge $e=(m,w)$ from $E_c$ because it is dominated by some newly added edge $(m',w)$. We want to show that $e$ cannot belong to any strongly stable matching  dominated by $M$. Suppose to the contrary that $e$ belongs to a strongly stable matching $N$ dominated by $M$. Since $rank_w(m,w) \neq rank_M(w)$, because $e \in E_c$, we get that $rank_N(w) \neq rank_M(w)$. Edge $e$ belongs  to $E_c$ and $m'$ has an incident edge in $E_c$. Therefore by Fact \ref{F} (3) $m'$ and $m$ belong to a common strongly connected component and hence $M(m') \succ_{m'} N(m')$. Then by Property \ref{pro2} $rank_{m'}(m',w) \leq rank_N(m')$. Therefore $e=(m,w)$ cannot belong to $N$ as it would be blocked by $(m',w)$. \koniec

\begin{lemma} \label{lem6}
Let $M'$ be a maximum matching in $G_c$,  $Z$  a set of men reachable from a free man $x_0$ by alternating paths and  $N(Z)$ women adjacent in $G_c$ to $Z$. Then assuming Algorithm satisfies Properties \ref{prop} and \ref{pro2}, edges of $G_c$ and lowest ranked edges of $E'$ incident to women in $N(Z)$ cannot be contained in any strongly stable matching dominated by $M$. 
\end{lemma}

%\subsection{Proof of Lemma \ref{lem6}} 
\dowod Let us assume that $e = (m, w)$ is an edge such that $w \in N(Z)$ and there is a strongly stable matching $N$ such that $e \in N$ and $M \prec N$.

We can easily prove that $|Z| = |N(Z)| + 1$ and that every woman in $N(Z)$ is matched in $M'$ with a man in $Z$. Edge $e$ is the lowest ranked edge incident to $w$.

Let $\tilde E$ be the set of edges incident to women in $N(Z)$ which Algorithm wants to remove. Consider $N \cap \tilde E$, let $U'$ be their female endpoints and $Z'$ be their male endpoints. From our assumptions it follows that $w \in U'$, hence $U' \neq \emptyset$. $N$ matches men in $Z'$ with women in $U'$, so $|Z'| = |U'| \leq |N(Z)| < |Z|$.

We will prove the existence of an edge $e' = (m', w')$ such that $m' \in Z \setminus Z'$ and $w' \in U'$ and then show that it blocks $N$.

Assume that $M'$ contains no such edge. Then it pairs women in $U'$ with men in $Z'$ and since $|U'| = |Z'|$, $M'$ pairs the men in $Z'$ with the women in $U'$. Hence $x_0 \in Z \setminus Z'$ as $x_0$ is free in $M'$. Consider the alternating path from $x_0$ to $w$. Let $(a, b)$ be the first edge on the path with $b \in U' \cup Z'$. If $b \in Z'$, $e'$ is a matching edge and $a \in U'$, contradicting the fact that $(a, b)$ is the first edge on the path with $b \in U' \cup Z'$. Thus $b \in U'$ and $a \in Z \setminus Z'$.

Since $w' \in U'$,  $rank_{w'}(N(w')) = rank_{w'}(m')$. We claim that in $N$ vertex $m'$ is either unmatched or matched to a woman strictly below $w'$ on his preference list. To prove this note that $m'$ cannot be matched to a vertex strictly better than $w'$ because $M$ dominates $N$ -- if follows easily from Property \ref{prop}. If $m'$ is matched in $N$ to a woman $w''$ such that $rank_{m'}(w'') = rank_{m'}(w')$, then $(m', w'') \in E_d$ from the definition of $G_d$. If $(m', w'') \in E_c$ then $m' \in Z'$, a contradiction. If $(m', w'') \notin E_c$, then at some point it must have been deleted from $E_c$ and by Lemma \ref{lem5} it cannot belong to $N$. Hence $m'$ has to be matched in $N$ to a vertex strictly worse than $w'$.

We conclude that $(m', w')$ blocks $N$, a contradiction. \koniec

\begin{lemma}
Suppose that matching $M_{i-1}$ output by Algorithm is strongly stable or $M_{i-1}=M_0$. Then 
matching $M_i$ output by Algorithm is strongly stable and is a strict successor of $M_{i-1}$.
\end{lemma}
\dowod By previous lemmas we can assume that at the moment of outputting $M_i$ Algorithm satisfies Properties \ref{prop} and \ref{pro2}.
$M_i$ is output because $M'$ is perfect on a strongly connected component $S$ with $outdeg(S)=0$. Thus $M_i$ is of the form $(M' \cap S) \cup (M\setminus S)$. 
First we prove that $M_i$ is strongly stable. Suppose to the contrary that $M_i$ is blocked by some edge $e=(m,w)$. We can notice that it cannot happen that exactly one of the vertices $m,w$ belongs to $S$. It is so because of the following. Suppose that $m \in S$. Then $e$ would be an outgoing edge of $S$, a contradiction. If $w \in S$ and $m \notin S$, then $M(m)=_m M_i(m)$ and $M(w)_w \prec M_i(w)$, which would mean that $e$ blocks $M=M_{i=1}$, a contradiction.

Hence, the endpoints of a blocking edge $e$ must both belong to $S$. Let us notice that at the moment of calculating alternating paths the edges of $E_c$ incident to vertex $v$ have the same rank with respect to $v$. Since $e$ blocks $M_i$,  it must have at some point belonged to $E_c$ and got deleted later. An edge incident to woman $w$ can get deleted only if it is dominated by another edge of $E_c$. This then means that the rank of edges currently incident to $w$, and thus to $M_i$, is higher than that of $e$ - a contradiction.

Now we prove that $M_i$ is a strict successor of $M_{i-1}$. Let  $v$ be any vertex of $S$ and $N$ any successor   of $M_{i-1}$ such that  $rank_N(v) \neq rank_{M_{i-1}}(v)$. Then  $rank_{M_{i-1}}(v) \neq rank_{M_i}(v)$. By Property \ref{prop1} the rank of every vertex of $S$ in $M_{i-1}$ must be different from its rank in  $N$.  By Property \ref{pro2}, for any  man $m$ of $S$ we have $rank_N(m) \geq rank_{M_i}(v)$. This concludes the proof. \koniec 

Finally, we have a lemma with an easy proof. 
\begin{lemma}
After the updating of graphs $G_c$ and $G_d$ in line $38$, Algorithm satisfies  Properties \ref{prop} and 
\ref{pro2}.
\end{lemma}

Using previous lemmas, we have proved:

\begin{theorem}
Algorithm computes a maximal sequence of strongly stable matchings.  
\end{theorem}

\subsection{Running Time of Algorithm} 

Without any additional modifications we can rather easily prove:

\begin{theorem} \label{czas1}
The running time of Algorithm  is $O(m^2)$
\end{theorem}
\dowod Each time we introduce a new edge or edges to $E_d$ we need to compute strongly connected components of $G_d$. Computing strongly connected components of any directed graph $G'=(V', E')$ can be done in $O(|E'|)$ time. Since each edge $e$ of $G$ is added to $G_d$ at most once and 
since $G_d$ is a subgraph of $G$ at all times, the overall time spent on computing strongly connected components of $G_d$ is $O(m^2)$.

Each time we introduce a new edge to $E_c$ we need to compute women reachable by alternating paths from free men in $G_c$. Each such computation takes $O(|E_c|)$ time. Every edge of $G$ is added to $G_c$ at most once and for all times $E_c \subseteq E$ - hence the time spent on computing alternating paths in $E_c$ during the whole execution of Algorithm is $O(m^2)$.

Updating graphs $G_d$ and $G_c$ takes $O(m)$ time overall. \koniec

Next, we show that Algorithm can be modified so that it runs in $O(nm)$ time. To this end we are going to use the concept of {\em levels} introduced in \cite{KMMP}. We define the {\em level} of an edge, vertex and matching in the same way as in \cite{KMMP}:

\begin{definition}
Let ${\cal E_i}$ be the edges added to $G_c$ in phase $i$ and define the {\em level} $l(e)$ of an edge $e$ to be the phae when this edge was first added to $G_c$. Edges never added to $G_c$ have no level assigned to them.
\end{definition} 

Thus, the set of edges ever added to $G_c$ consists of the disjoint union $ {\cal E_1 \cup E_2 \cup \ldots E_r}$, where $r$ is the total number of phases in the algorithm. Note that $r \leq m$.

\begin{definition}
Define the {\em level} $l(v)$ of a vertex $v$ to be the minimum level of the edges in $G_c$ incident to $v$. The level of an isolated vertex is undefined.
\end{definition}

\begin{definition}
The {\em level} $l(M)$ of a matching $M$ is the sum of the levels of the matched women. A matching $M$ is level-maximal if $l(M) \geq l(M')$ for any matching $M'$ which matches the same men.
\end{definition}

We show that in order to make Algorithm  run in $O(nm)$ time it suffices to change Line $24$. Line $24$ of the modified algorithm, called Algorithm  Mod, is: "let $w$ be a free woman in $G_c$ {\em of maximal level} reachable from $m$ by an alternating path $p$ in $E_c$". Because of this modification we prove:

\begin{lemma}
Matching $M'$ is level-maximal at all times of the execution of Algorithm  Mod.
\end{lemma}

The proof of this lemma is the same as that of its analogue in \cite{KMMP} and is based on the following lemmas, also proved in \cite{KMMP}:

\begin{lemma}
For a man, all incident edges in $G_c$ have the same level. All women adjacent to a man of level $i$ have level at most $i$. When a woman loses an incident edge in $E_c$ she loses all her incident edges in $E_c$.
\end{lemma}

\begin{lemma} 
A matching $M$ is level-maximal iff there is no alternating path in $G_c$ from a free woman in $M'$ to a woman of lower level.
\end{lemma}

\begin{lemma}
If $M'$ is level-maximal, $m$ is a free man in $M'$, $w$ is a woman of maximal level reachable from $m$ by an augmenting path $p$, then $N=M' \oplus p$ is level-maximal.
\end{lemma}

The search for augmenting paths in $G_c$ and its analysis are also the same as in \cite{KMMP}.

Therefore we have:

\begin{lemma}\label{paths}
The total time of Algorithm  Mod spent on computing augmenting paths in $G_c$, i.e. on lines $22 - 31$, is bounded by $O(nm)$.
\end{lemma}

Below we show that the time needed to compute strongly connected components during the execution of Algorithm  Mod can be estimated more carefully than in Theorem \ref{czas1}.

\begin{lemma} \label{comp}
The overall time of Algorithm  Mod spent on computing strongly connected components of $G_d$ is $O(nm)$.
\end{lemma}
\dowod Pearce \cite{Pearce} and Pearce and Kelly \cite{PK} sketch how to extend their algorithm and that of Marchetti-Spaccamela et al. \cite{MarchettiNR96}
to strong component maintenance. Their algorithm runs in $O(nm)$ time if edges can only be added to the graph and not deleted and $n, m$ denote the number of  vertices and edges, respectively. In Algorithm (and Algorithm Mod) edges of $G_d$ can be deleted -- in line 38. However,
they are deleted only when $M'$ is perfect on a strongly connected component $S$. As a result only a strongly connected component $S$ of $G_d$
vanishes and other strongly connected components are unaffected. Some of the edges of $G_d$ having one or two endpoints in $S$ remain in $G_d$.
We can treat them as though they were added anew to the graph. Since the ranks of men increase as we output subsequent strongly stable matchings, we can notice that each edge can be added anew to the graph $G_d$ at most three times. This proves the lemma. \koniec

As a consequence of Lemmas \ref{paths} and \ref{comp} we obtain:
\begin{theorem}
Algorithm Mod runs in $O(nm)$ time.
\end{theorem}

\section{Rotations}
Based on a maximal sequence ${\cal C}$ of strongly stable matchings it is possible to build a concise representation of the set of all strongly stable matchings. It is done very similarly as in the classical stable matching problem without ties. There such a representation is constructed from an analogous  maximal sequence ${\cal D}$ of stable matchings $M'_0 \succ M'_1 \succ \ldots \succ M'_z$, where $M'_0$ and $M'_z$ denote appropriately a man-optimal and woman-optimal stable matching. Let us note that in both problems a maximal sequence of (strongly) stable matchings is not unique. 

The symmetric difference $M \oplus N$ of two matchings, with the same sets of matched vertices consists of alternating cycles.
A symmetric difference $M_{i-1}\oplus M_i$ of two consecutive stable matchings in ${\cal D}$ is called a {\em rotation}. It turns out that in the case of the stable matching problem without ties every rotation consists of  one alternating cycle and irrespective of a maximal sequence ${\cal D}$ of stable matchings one always gets the same set of rotations. The set of stable matchings is characterised by a partial order $(\Pi, \leq)$  on rotations with a relation of preceding defined as follows.  We say that rotation $R_1$ precedes rotation $R_2$ and denote $R_1 \leq R_2$ if in every maximal sequences ${\cal D}$  of a given instance a rotation $R_1$ occurs before a rotation $R_2$. Every stable matching corresponds to a closed subset of $\Pi$.  Theory regarding rotations is very well described in the book by Gusfield and Iriving \cite{GI}.

In the case of strongly stable matchings we proceed analogously. We define a {rotation} as a symmetric difference $M_{i-1}\oplus M_i$ of two consecutive strongly stable matchings in ${\cal D}$. This time, however, a rotation may consist of more than one alternating cycle. Also, we define an equivalence class on rotations so that $R_1=M\oplus N$ is equivalent to $R_2=M'\oplus N'$ if and only if $M \sim M'$ and $N \sim N'$. On the set of classes of rotations we construct a partial order $(\Pi', {\cal \leq'})$ in time $O(nm)$.  

\bibliographystyle{abbrv}
\bibliography{bibliography}

\end{document}